\documentstyle[aps]{revtex}
\newcommand{\bba}{\begin{eqnarray}}
\newcommand{\eea}{\end{eqnarray}}
\newcommand{\bb}{\begin{equation}}
\newcommand{\ee}{\end{equation}}
\newcommand{\bban}{\begin{eqnarray*}}
\newcommand{\eean}{\end{eqnarray*}}

\def\a{\alpha}
\def\b{\beta}
\def\g{\gamma}
\def\c{\chi}
\def\d{\delta}

\def\f{\phi}

\def\m{\mu}
\def\n{\nu}

\def\r{\rho}
\def\s{\sigma}

\def\x{\xi}

\def\G{\Gamma}

\def\O{\Omega}

\title{\hfill{\normalsize BROWN-HET-1219, DAMTP-2000-41, OUTP-00-18P \\[0.3cm]}
       Cosmological Perturbations in Brane-World Theories: Formalism}
\author{Carsten van de Bruck$^{1,2}$\footnote{email: C.VanDeBruck@damtp.cam.ac.uk},
Miquel Dorca$^{1}$\footnote{email: dorca@het.brown.edu}, 
Robert H. Brandenberger$^{1}$\footnote{email: rhb@het.brown.edu},
Andr\'e Lukas$^{3}$\footnote{email: lukas@thphys.ox.ac.uk} \\
$^{1}${\small Department of Physics, Brown University, Providence, RI 02912, USA}\\
$^{2}${\small DAMTP, Center for Mathematical Sciences, Wilberforce
       Road, CB3 0WA, Cambridge, United Kingdom}\\
$^{3}${\small Department of Physics, Theoretical Physics, University of Oxford}\\
{\small 1 Keble Road, Oxford OX1 3NP, United Kingdom}}
\begin{document}
\noindent
\maketitle
\begin{abstract}
We develop a gauge-invariant formalism to describe metric 
perturbations in five-dimensional brane-world 
theories. In particular, this formalism applies to models
originating from heterotic M-theory. We introduce a generalized 
longitudinal gauge for scalar perturbations. 
As an application, we discuss some aspects of the evolution
of fluctuations on the brane. Moreover, we show how the
five-dimensional formalism can be matched to the known
four-dimensional one in the limit where an effective four-dimensional
description is appropriate.
\end{abstract}

\section{Introduction}

In recent years, the way in which string-theory is believed to
be connected to observable physics has changed dramatically.
The new viewpoint is mainly due to two ideas, namely the brane-world
idea~[\ref{Horava}]--[\ref{kt}] and the idea that couplings and scales
of additional dimensions are much more flexible than previously
assumed~[\ref{witten},\ref{lyk},\ref{Dvali1},\ref{Dvali2}].
Not only have these ideas led to new directions in M-theory
phenomenology and, more generally, string-theory inspired particle
phenomenology, but also in early universe cosmology.

Much of the recent activity in brane-world cosmology is centered
around five-dimensional brane-world theories, see for
example~[\ref{lowcosm}]--[\ref{viss2}].
A large class of such theories arises from heterotic
M-theory~[\ref{losw1},\ref{elpp},\ref{losw2}]. Other five-dimensional
models have been introduced in ref.~[\ref{visser}]--[\ref{lykken}]
which may provide an alternative solution to the hierarchy problem.

\medskip

A central question is whether the possible existence of a brane-world
and large additional dimensions in the early universe leads to
observable consequences today. Specifically, cosmological perturbations
as, for example, observed in the cosmic microwave background provide us
with a window to the early universe that, perhaps, can be used to test
the brane-world idea. It is with this motivation in mind, that we
set out to study metric perturbations in brane-world models.
It may not be immediately clear that the existence of additional
dimensions and branes should have important consequences for the
formation and evolution of cosmological perturbations. Let us, as a
comparison, consider ``traditional'' Kaluza-Klein cosmologies, where
the higher-dimensional universe is usually split into a product of
maximally symmetric subspaces each one with an individual scale
factor. Cosmological perturbations are normally treated at the
linearized level and, hence, in such Kaluza-Klein
cosmologies~[\ref{fabris}]--[\ref{abbott}] internal and external fluctuations
basically decouple. The additional dimensions effect perturbations of
the three-dimensional universe only via the kinematics of the
background scale factors (and/or dilaton background fields
[\ref{dilaton}]). The situation is quite different for
brane-world theories. The branes are stretched across the
three-dimensional universe and are located at specific points in the
internal space. Moreover, they carry worldvolume fields that can
only propagate on the brane and that are likely to be excited in the
early universe, both coherently and thermally. As a consequence, the
cosmological background in an early brane-world universe is highly
inhomogeneous in the additional dimensions since the branes constitute
localized sources of stress-energy. Even if perturbations
around such a background are treated at the linearized level,
the perturbations of the three-dimensional universe are effected by
the non-linear distortion of the cosmological background in the
internal dimensions. This constitutes a crucial difference between
conventional Kaluza-Klein cosmology and brane-world cosmology which is
directly related to the presence of branes. It is this difference
that may lead to new predictions for cosmological perturbations in
brane-world models and that motivates the present investigation.

\medskip

In this paper, we will develop a formalism for metric fluctuations in
brane-world theories that takes the characteristic property of
brane-world cosmologies, the above mentioned inhomogeneity in the
additional dimensions, into account. We understand such a formal
development as a first step towards analyzing predictions for
cosmological perturbations in brane-world theories. In the next
section, we start out to generalize the well-known formalism of
four-dimensional gauge-invariant metric perturbations~[\ref{Bardeen}],
[\ref{Brandenberger}] to brane-world theories with an arbitrary
number of additional dimensions. Subsequently, in section III, we
focus on five-dimensional brane-world models on the orbifold $S^1/Z_2$
related to those originating from heterotic
M-theory~[\ref{losw1},\ref{elpp},\ref{losw2}]. Specifically, we
consider the five-dimensional Einstein equation coupled to bulk as
well as brane stress-energy. For such a theory we derive the equations
of motion for scalar metric perturbations in a generalized
longitudinal gauge. Consistency of these equations is used to
determine the most general structure of the stress-energy on the
brane. In section IV, these results are applied to find the Israel
matching conditions~[\ref{Israel}] for the scalar metric
perturbations restricted to the branes. Finally, in section V, we show
how our formalism for brane-world metric perturbations is related to
the conventional one in four-dimensions. This is done by a matching
procedure applied in a limit where the five-dimensional brane-world theory
has an effective four-dimensional description.

\section{Gauge-invariant variables}

In this section, we will develop a gauge-invariant formalism for
metric perturbations in brane--world models. Using such a
gauge-invariant approach is particularly useful in order to identify
the correct physical degrees of freedom. Once this has been done, a
specific gauge can be chosen in order to simplify the subsequent
equations. Specifically, we will use a generalized longitudinal gauge
later on. However, as a warm-up for the higher--dimensional case,
we would first like to review the well-known four-dimensional
gauge-invariant formalism following ref.~[\ref{Brandenberger}].

\subsection{The four-dimensional formalism}

Starting point is a background metric with a maximally symmetric
three--dimensional spatial space. This metric is of the form
\begin{equation}
 ds_4^2 = a_4^2\left\{dt^2-\Omega_{ij}dx^idx^j\right\} \label{bm4}\;,
\end{equation}
where indices $i,j,\cdots = 1,2,3$ run over the three spatial indices.
Indices $\m ,\n ,\cdots = 0,1,2,3$ are used to index four-dimensional space-time.
Furthermore, $a_4=a_4(t)$ is the four-dimensional scale factor and $\O_{ij}$
is the metric of the three--dimensional maximally symmetric space
explicitly given by
\bb\label{Omega}
\Omega_{ij}=\delta _{ij}\left[1+{1\over 4}kx^lx^m\d_{lm}\right]^{-2}\; .
\ee
Here $k=0,1,-1$ corresponds to a flat, closed or open universe,
respectively. The idea is now to classify perturbations of the
metric~(\ref{bm4}) according to their transformation properties with
respect to the maximally symmetric space. This leads to the perturbed
metric
\begin{equation}
 ds_4^2 = a_4^2\left\{ (1+2\phi_4 )dt^2-\left[
          (1-2\psi_4)\O_{ij}+2E_{4|ij}+2F_{4(i|j)}+h_{4ij}\right]dx^idx^j
          +W_{4i}dtdx^i\right\}\;.\label{pert4}
\end{equation}
Here and in the following four-dimensional quantities are indexed
by ``$4$'' to distinguish them from their higher-dimensional
counterparts to be introduced later. The vertical bar refers to a
covariant derivative with respect to the metric $\O_{ij}$.
The vector $F_{4i}$ has a vanishing divergence, that is
${F_{4i}}^{|i}=0$ and the tensor $h_{4ij}$ is traceless and
divergence-less, that is ${h_{4i}}^i=0$ and ${h_{4i}^j}_{|j}=0$.
In addition, we can decompose the off-diagonal perturbation $W_{4i}$
further into the gradient of a scalar $B_4$ and a divergence-less
vector $S_{4i}$. Explicitly, this reads
\bb
 W_{4i} = B_{4|i}+S_{4i}\; .
\ee
Consequently, we have four scalar metric perturbations $(\f_4 ,\psi_4
,E_4,B_4)$, two vector perturbations $(F_{4i}, S_{4i})$ and a tensor
perturbation $(h_{4ij})$. All these perturbations are functions of time
as well as of the spatial coordinates $x^i$, of course. Next we consider
an infinitesimal coordinate transformation
\bb
 x^\m\rightarrow\tilde{x}^\m = x^\m+\x^\m \;,
\ee
where the vector $\x^{\mu}$ depends on all four coordinates, in general.
The corresponding infinitesimal change of the metric is given by
\bb
 g_{4\m\n}\rightarrow\tilde{g}_{4\m\n}=g_{4\m\n}-2\nabla_{(\m}\x_{\n )}\; .
 \label{t4}
\ee
To understand how this coordinate transformation acts on the metric
perturbations we split $\x^\m$ as $\x^\m =(\x^0,\x^i)$ into a time and
a spatial part. The spatial component $\x^i$ can be decomposed further
into a gradient and a divergence-less part as
\bb\label{etai}
\xi ^i=\xi ^{|i}+\eta ^i\; .
\ee
As a result, the transformation parameter $\x^\m$ contains two scalar
components $(\x^0,\x)$ and one vector component $(\x^i)$. Given this
setup, one can compute the transformation properties of the metric
perturbations by applying eq.~(\ref{t4}) to the perturbed
metric~(\ref{pert4}) and taking into account that
$\x_\m =a_4^2\, (\x_0,-\x_i)$. For the scalar perturbations one finds
\bba
 \tilde{\f}_4 &=&\f -H_4\x^0-\dot{\x}^0\;, \\
 \tilde{\psi}_4 &=&\psi_4+H_4\x^0\;, \\
 \tilde{B}_4 &=& B_4+\x^0-\dot{\x}\;, \\
 \tilde{E}_4 &=& E_4-\x\; .
\eea
Here, $H_4$ is the Hubble parameter defined by $H_4=\dot{a}_4/a_4$.
The vector perturbations transform as
\bba
 \tilde{F}_{4i} &=& F_{4i}-\eta_i\;, \\
 \tilde{S}_{4i} &=& S_{4i}-\dot{\eta}_i\;,
\eea
while the tensor perturbation $h_{4ij}$ is invariant. In these
equations, spatial indices are lowered and raised with the metric
$\O_{ij}$, that is, for example $\xi _{|i}=\Omega _{ij}\xi ^{|j}$ and
$\eta _i=\Omega _{ij}\eta ^j$. With these results, it is
straightforward to introduce the following gauge-invariant variables.
\begin{center}
\underline{{\it Scalar variables}}
\end{center}
\begin{eqnarray}\label{gaugeinvsca4}
\qquad\quad\Phi_4 &=&\f_4+H_4(B_4-\dot{E}_4)+
                     \dot{B}_4-\ddot{E}_4 \label{sca4}\\
\qquad\quad\Psi_4 &=& \psi_4 - H_4(B_4 - \dot{E}_4).\label{sca41}
\end{eqnarray}
\begin{center}
\underline{{\it Vector variables}}
\end{center}
\begin{eqnarray}\label{gaugeinvvec4}
\qquad\quad{\cal F}_{i} = S_{4i} - \dot{F}_{4i}\;.
\end{eqnarray}
\begin{center}
\underline{{\it Tensor variables}}
\end{center}
\begin{equation}\label{gaugeinvten4}
\qquad h_{4ij}\;.
\end{equation}
As the physical degrees of freedom, one has therefore identified two
scalar perturbations, one vector perturbation and one tensor
perturbation. Of particular importance are the two scalar
perturbations $\Phi_4$ and $\Psi_4$ on which we are going to focus.
The expressions~(\ref{sca4}) and (\ref{sca41}) for these perturbations suggest
the gauge choice $B_4=E_4=0$ in the scalar sector which is referred to as
{\em longitudinal gauge}. Clearly, from the above transformation properties
of the scalar perturbations such a choice can be made. Then, the
gauge-invariant scalar variables coincide with the ``original'' variables,
that is, $\Phi_4=\f_4$ and $\Psi_4=\psi_4$. This gauge choice considerably
simplifies subsequent calculations and its generalization will be
quite helpful to deal with the higher-dimensional case. The perturbed
metric then takes the form
\bb
 ds_4^2 = a_4^2\left\{ (1+2\f_4)dt^2-(1-2\psi_4)\O_{ij}dx^idx^j\right\}\;.
 \label{stp4}
\ee
Finally, we need to specify the stress-energy. For the background,
by the maximal symmetry of the three-dimensional spatial space, it is
dictated to be of the form
\bb
 {{T_4}^\m}_\n = \text{diag}(\r_4,-p_4,-p_4,-p_4)\; .
 \label{T4}
\ee
The scalar perturbations to this stress-energy can be written as
\bb
 \d{{T_4}^\m}_\n = \left(\begin{array}{cc}
                   \d\r_4&-(\r_4+p_4)a_4^{-1}v_{4|j}\\
                   (\r_4+p_4)a_4{v_4}^{|i}&-
                   \d p_4{\d^i}_j+{{\s_4}^{|i}}_{|j}
                   \end{array}\right)
 \label{dT4}
\ee
with the potential $v_4$ for the velocity field $v_{4|i}$ and the
quantity $\s$ specifying the anisotropic stress. The equations of
motion for the background and the scalar perturbations subject to the
above stress-energy are given in ref.~[\ref{Brandenberger}] and will
not be repeated here. These equations form the basis for the study of
cosmological perturbations and we now turn to develop their
higher-dimensional generalization.

\subsection{Gauge-invariant variables in brane-world theories}

We would now like to proceed in close analogy with the
four-dimensional case reviewed above and develop a gauge-invariant
formalism of metric perturbations in brane-world theories. First, we
consider the general situation of $d$ additional dimensions although
later we will be more specific and focus on the case $d=1$, that is, a
five-dimensional universe. The coordinates of the additional
dimensions are denoted by $(y^5,\cdots ,y^{4+d})$. For the purpose of
this subsection, all we need to specify is that the branes are
stretched across the usual four-dimensional space-time and are located
at specific points (or submanifolds) in the additional dimensions. We
will be more precise about the brane positions later when we consider the
five-dimensional case.

How should metric perturbations be classified in such a brane-world
theory? In the previous four-dimensional case we have used their
tensor properties with respect to the three-dimensional spatial
subspace for this classification. At first glance, one might now want
to use their tensor properties with respect to the $3+d$-dimensional
spatial space. The cosmological principle, of course, only asserts the
maximal symmetry of the usual three-dimensional space but the maximal
symmetry of the $d$-dimensional internal space may be taken as an
additional, simplifying assumption. It is at this point, that the
brane-world nature of the theory comes into the game. Since the branes
are localized in the additional dimensions the assumption of maximal
symmetry cannot, in general, be extended to those dimensions. In fact,
as will become more explicit below, the branes lead to stress-energy
in the Einstein equation localized in the additional dimensions and,
hence, the symmetry of the background metric will typically not be
enhanced with respect to the four-dimensional case. Consequently, we
split the coordinates into two groups, namely the
inhomogeneous coordinates $(y^a)=(t,y^5,\cdots ,y^{4+d})$ on which
the background metric generally depends in a non-trivial way and
the usual three spatial coordinates $(x^i)$ corresponding to the
maximally symmetric space. In the following we use
indices $a,b,\cdots = 0,5,\cdots ,4+d$ for time and the additional
dimensions, indices $i,j,\cdots =1,2,3$ for the three--dimensional
space and indices $\a ,\b ,\cdots = 0,1,2,3,5,\cdots ,4+d$ for the
full $4+d$-dimensional space-time. Then the most general
higher-dimensional metric consistent with the maximally symmetric
three-dimensional spatial manifold is given by
\begin{equation}
 ds^2 = a^2\left\{\gamma_{ab}dy^ady^b-\Omega_{ij}dx^idx^j\right\},
\end{equation}                   
where the scale factor $a$ and the metric $\g_{ab}$ are functions of
the coordinates $y^a$ only. Here $\Omega_{ij}$ is the metric
of the maximally symmetric space of constant curvature given in (\ref{Omega}). 
Given this structure of the background metric, we are forced to
classify metric perturbations by their three-dimensional tensor
properties as in the four-dimensional case. We stress again that this
is a direct consequence of the brane-world nature of the theory that
we are considering. With these remarks in mind, the higher-dimensional
generalization of the perturbed metric~(\ref{pert4}) can be written in
the form
\begin{equation}\label{permetric1}
ds^2 =  a^2\left\{\gamma_{ac}\left( \delta^{c}_{b} +
        2\phi^{c}_{b}\right) dy^a dy^b 
       -\left[ \left(1 - 2\psi \right) \Omega_{ij} + 2E_{|ij} +
       2F_{(i|j)} + h_{ij}\right] dx^i dx^j-2 W_{ai}dy^a dx^i\right\}\; .
\end{equation}
As in the four-dimensional case, $F_i$ and $h_{ij}$ have a vanishing
divergence and, in addition, $h_{ij}$ is traceless. 
As before, the three-vectors $W_{ai}$ can be split as follows, 
\bb
W_{ai} = B_{a|i} + S_{ai}\; ,
\ee
where ${{S_a}^i}_{|i}=0$. Observe that the perturbed metric
(\ref{permetric1}), defined in this way, is completely general. In
fact, this can be easily seen by counting degrees of freedom. As an
example, we can consider the simplest case of only one extra dimension
setting $y=y^5$ and $a,b,\cdots =0,5$. Then, the most general perturbed metric
contains $15$ degrees of freedom, which are parameterized by the seven
scalar perturbations ($\phi _0^0$, $\phi _5^0$, $\phi _5^5$, $\psi$,
$E$, $B_a$), six degrees of freedom from the vector perturbations ($F_i$,
$S_{ai}$) and two degrees of freedom from the tensor perturbation
$h_{ij}$. Of course, in counting the degrees of freedom originating from
vector and tensor perturbations we have taken the constraints on these
quantities into account.

Let us now return to the general case of $d$ additional dimensions and
consider the coordinate transformations
\begin{eqnarray}\label{gaugetransf}
x^{\a} \rightarrow \tilde{x^\a} = x^{\a} + \xi^{\a} 
\end{eqnarray}
with infinitesimal parameters $\x^\a$. In accordance with the above
discussion, we split these parameters as $(\x^\a ) = (\x^a,\x^i)$.
We adopt the useful convention that indices of type $a$ ($i$) are
lowered, raised and contracted using the metric $\g_{ab}$ ($\O_{ij}$).
Furthermore, we take the vertical bar to denote the covariant
derivative with respect to $\g_{ab}$ or $\O_{ij}$ depending on the
index type. From the transformation law 
\begin{equation}
{\tilde g}_{\a\b} =  {g}_{\a\b} - 2\nabla_{(\a} \xi_{\b )},\label{t5}
\end{equation}
of the metric, and taking into account that $\x_\a  =a^2\, (\x_a,-\x_i)$,  
we find for the transformation of the scalar perturbations
\begin{eqnarray}\label{gaugetransfg}
\delta \phi_{ab} &=& -\xi_{(a|b)}-H^c\x_c\g_{ab},\label{gaugetransfg1}\\
\delta \psi &=& H^a \xi_a,\\
\delta E &=& -\xi,\label{gaugetransfg3}\\
\delta B_a &=& \xi _{a} - \xi _{|a}\label{gaugetransfg4}.
\end{eqnarray}
where we have introduced the generalized Hubble parameters
\bb
 H_c=\frac{a_{|c}}{a}\; .
\ee
The vector perturbations in the metric (\ref{permetric1}) change according to
\begin{eqnarray}
\delta F_i &=& -\eta_i, \\
\delta S_{ai} &=& -\eta_{i|a}\; .
\end{eqnarray}
Finally, the tensor perturbation $h_{ij}$ is invariant under 
the first order gauge transformation (\ref{t5}). From these results
we easily find the following gauge-invariant quantities.
\begin{center}
\underline{{\it Scalar variables}}
\end{center}
\begin{eqnarray}\label{gaugeinvsca}
\qquad\quad\Phi_{ab} &=& \phi_{ab} + H^c(B_c-E_{|c})\g_{ab}+
                     (B_{(a}-E_{|(a})_{|b)}\\
\qquad\quad\Psi &=& \psi - H^c \left(B_c - E_{|c}\right).
\end{eqnarray}
\begin{center}
\underline{{\it Vector variables}}
\end{center}
\begin{eqnarray}\label{gaugeinvvec}
\qquad\quad {\cal F}_{ai} = S_{ai} - F_{i|a}\;.
\end{eqnarray}
\begin{center}
\underline{{\it Tensor variables}}
\end{center}
\begin{equation}\label{gaugeinvten}
\qquad h_{ij}\;.
\end{equation}
We conclude that the physical degrees of freedom consist of the
$(d+1)(d+2)/2+1$ gauge invariant scalar perturbations $(\Phi_{ab},\Psi )$,
$d+1$ gauge invariant vector perturbations ${\cal F}_{ai}$ and a
gauge invariant tensor perturbation $h_{ij}$. The above
gauge-invariant variables are a direct generalization of the
corresponding four-dimensional ones. Specifically, restricting to no
additional dimensions and setting $\g_{00}=1$,
eqs.~(\ref{gaugeinvsca})--(\ref{gaugeinvten}) exactly reproduce the
four-dimensional expressions (\ref{gaugeinvsca4})--(\ref{gaugeinvten4}).
However, in the case $d>0$ our formalism clearly has a richer
structure than the conventional four-dimensional one. 

\subsection{A generalized longitudinal gauge for scalar perturbations}

In the subsequent sections we will focus on the 
evolution of scalar perturbations. Vector and tensor perturbations
will be discussed elsewhere. In order to simplify this
discussion we introduce a {\em generalized longitudinal gauge} for the
scalar perturbations. In analogy with the four-dimensional case, this
gauge is specified by
\bb
 B_a=0\; ,\qquad E=0\label{lg}\; .
\ee
Setting these quantities to zero can indeed be achieved by an
appropriate choice of the scalar transformation parameters $\x_a$
and $\x$ in the eqs.~(\ref{gaugetransfg1})--(\ref{gaugetransfg4}).
Note that we have exactly the correct number of transformation
parameters to do this and that, consequently, the gauge ambiguity
in the scalar sector is complete eliminated by this choice. Then,
the scalar part of the metric takes the simple form
\bb
 ds^2 = a^2\left\{\g_{ac}(\d_b^c+2\f_b^c)dy^ady^b-(1-2\psi )
        \O_{ij}dx^idx^j\right\}\; .
\ee
Furthermore, in this gauge, the scalar perturbations $\f_{ab}$ and
$\psi$ coincide with their gauge-invariant counterparts, that
is
\bb
 \Phi_{ab}=\f_{ab}\; ,\qquad \Psi = \psi
\ee
as it is the case in four dimensions.

\subsection{The five-dimensional case}

Let us restrict in this section and for the rest of the paper to the
case of a single extra dimension $y=y^5$. Then, the indices
$a,b,\cdots$ run over the values $0,5$ only. Furthermore, in order to
be more explicit, we choose the conformal gauge
\bb
 (\g_{ab})=b^2\text{diag}(1,-1)
\ee
for the background metric $\g_{ab}$ by performing a large gauge
transformation. Here $b=b(t,y)$ is a new, independent scale factor.
Then, the perturbed five-dimensional metric (\ref{permetric1})
reduces to
\bba
ds^2 &=& a^2\left\{ b^2\left[(1+2\phi)dt^2 -2Wdtdy 
       - (1-2\Gamma) dy^2 \right]\right.\nonumber\\
     &&\qquad\left. - \left[ \Omega _{ij}(1 - 2\psi) + 2 E_{|ij} + 2 F_{(i|j)} 
       + h_{ij} \right] dx^i dx^j-2W_{0i}dt dx^i-2W_{5i}dydx^i\right\}\;,
\eea
where we have defined
\bb
 \f = \f_0^0\; ,\qquad \G = -\f_5^5\; ,\qquad W=2\f_0^5=-2\f_5^0\; .
\ee
Recall that the scale factors $a$ and $b$ are functions of the
coordinates $t$ and $y$ only while the perturbations depend on all 
spacetime coordinates. The scalar gauge-invariant variables defined in
eq.~(\ref{gaugeinvsca}) can now be written more explicitly as
\bba
 \Phi_1 &\equiv&\Phi_0^0 = 
 \f + \frac{1}{b^2} \left[ (H_0-{\cal H}_0)(B_0-\dot{E})+\dot{B}_0-\ddot{E}
-(H_5+{\cal H}_5)(B_5 - E')\right]\\
 \Phi_2 &\equiv&\Phi_5^5 = \G -\frac{1}{b^2}\left[ 
(H_5-{\cal H}_5)(B_5-E')+ B_5'- E''- (H_0+{\cal
H}_0)(B_0-\dot{E})\right] \\
 \Phi_3 &\equiv&\Phi_0^5 =
        \frac{W}{2}-\frac{1}{2b^2}\left[(B_0'+\dot{B}_5)-2\dot{E}'
        -2{\cal H}_5(B_0 - \dot E) - 2{\cal H}_0(B_5 - E')\right]\\
 \Phi_4 &\equiv&\Psi = \psi -\frac{1}{b^2}\left[ H_0(B_0-\dot{E})
        -H_5(B_5-E')\right]\; .
\eea
Here and in the following the dot (prime) denotes the derivative with respect
to time (the coordinate $y$). Furthermore, we have introduced a second
set of ``Hubble''--parameters ${\cal H}_a = b_{|a}/b$.
Let us specialize these results
to the generalized longitudinal gauge defined by $B_0=B_5=E=0$.
Then the above scalar gauge-invariant variables coincide with $\f$,
$\psi$, $\G$ and $W$. The metric simplifies to
\bb
 ds^2 = a^2\left\{ b^2\left[(1+2\f )dt^2-2Wdydt-(1-2\G )dy^2\right]
        -(1-2\psi )\O_{ij}dx^idx^j\right\}\; .\label{stp}
\ee
This metric will be the starting point for our treatment of scalar
perturbations in the following sections. In addition to the
perturbations $\phi$ and $\psi$ that we are familiar with from the
four-dimensional case it contains two new perturbations, $\G$ and $W$,
that are related to the presence of the fifth dimension. 

\section{The perturbed Einstein equation in the longitudinal gauge}

As we have previously mentioned, the main application we have in mind
for this paper is a compactification of a five-dimensional theory
on the orbifold $S_1 / Z_2$. We start by compactifying the fifth
dimension on a circle restricting the corresponding coordinate $y$ to
the range $y\in [-R,R]$ with the endpoints identified. The action of
the $Z_2$ orbifolding symmetry on the circle is taken to be
$y\rightarrow -y$. Consequently, there exist two fix points at
$y=y_1=0$ and $y=y_2=R$. We assume that the three-branes, stretching
across $3+1$--dimensional space-time, are located at these fix points
in the orbifold direction. This setup is appropriate for a large
class of five-dimensional heterotic M-theory
models~[\ref{losw1},\ref{elpp},\ref{losw2}] that originate from 11--dimensional
Ho\v rava-Witten theory. It also applies to the five-dimensional
models introduced in ref.~[\ref{RS1},\ref{RS2}].

Next, we should truncate the five-dimensional metric in order to make
it consistent with the orbifolding. Since the metric has to be
intrinsically even under the $Z_2$ action its various components
satisfy the constraints
\begin{eqnarray}
g_{\mu\nu}(-y) &=& g_{\mu\nu}(y),\\
g_{\mu 5}(-y) &=& -g_{\mu 5}(y),\\
g_{55}(-y) &=& g_{55}(y).
\end{eqnarray}
At the same time, we have to make sure that coordinate transformations
do not lead out of the class of metrics defined this way. The
parameter $\x^\a$ for an infinitesimal coordinate transformation has,
therefore, to be restricted by
\bba
 \x^\m (-y) &=& \x^\m (y)\;, \\
 \x^5 (-y) &=& -\x^5(y)\; ,
\eea
which directly follows from~(\ref{t5}). From these rules we can 
deduce the $Z_2$ properties of the various
quantities in metric~(\ref{stp}) for scalar perturbations. While the
background scale factors $a$, $b$ as well as the perturbations $\phi$,
$\psi$ and $\G$ are $Z_2$ even, that is, for example, $a(-y)=a(y)$,
the perturbation $W$ is $Z_2$ odd, that is $W(-y)=-W(y)$. Similarly,
for the scalar components in the transformation parameter $\x^\a$,
we find that $\x_0$ and $\x$ are
even while $\x_5$ is odd. Also note that the derivative along the
fifth dimension of an odd variable is even and vice versa. For
instance, $W'(y) = W'(-y)$. Based on these $Z_2$ truncations we should
now discuss the continuity properties of all quantities. Normally,
one requires the metric to be continuous across the whole of
space-time in order to have a sensible notion of length and time. We
will also adopt this viewpoint, however with an additional subtlety.
Since the orbifolding identifies the upper and lower half of the
circle in the fifth dimension already one of them, say
the upper half, constitutes the whole of space-time. In fact, instead
of working with the {\em orbifold picture} where one keeps the full circle
as we do here, one could also use the {\em boundary picture} where only
one half of the circle (a line-segment) is considered. This shows that
a jump of a metric component at an orbifold fix point does not
contradict the continuity requirement. Of course, such a jump is
possible only for an odd component of the metric. Concretely, we
therefore require that all components of the metric~(\ref{stp}) are
continuous across the full orbicircle except for the odd component $W$
which may jump at the fix points (but is continuous otherwise). 
Corresponding assumptions have to be made for the parameter
$\x^\a$ so that coordinate transformations do not change these continuity
properties of the metric. Clearly the even components $\x_0$ and $\x$ have
to be continuous then. Is the odd component $\x_5$ allowed to jump
at the orbifold points? Eq.~(\ref{gaugetransfg1}) shows that
$\G = -\f^5_5$, which has to be even and continuous, transforms with
the derivative $\x_5'$. Hence, if $\x_5$ jumped at the fix points it
would lead to a delta-function singularity in the metric which is 
clearly unacceptable. We, therefore, have to require that $\x_5$
is continuous everywhere on the orbicircle\footnote{From this 
conclusion we see, that we have glossed over a subtlety when 
introducing the generalized longitudinal gauge. Clearly, 
for continous $\x_5$ the quantity $B_5$ can only be gauged to 
zero if $E_{|5} - B_5$ is continous, as can be seen from 
(\ref{gaugetransfg3}) and (\ref{gaugetransfg4}). 
We will, therefore, in addition require the 
continuity of $E_{|5} - B_5$.}. In particular, this means
that $\x_5$ vanishes at the fix points, that is $\x_5(y_n)=0$.

It is clear that the above conclusions depend somewhat on the fact
that we are working with an orbifold. For example, if we had
considered compactification on a circle instead,
all components of the metric had to be continuous. Correspondingly,
some of the conclusions below will be slightly modified for other
compactifications, however, in a way that is usually rather obvious.

\medskip

Given this setup the Einstein equations can be written as
\begin{equation}\label{fieldeq}
G_{\a\b}\equiv R_{\a\b} - \frac{1}{2}g_{\a\b} R = T_{\a\b} 
+ \sum_{n=1}^{2}T^{(n)}_{\a\b} \delta(y-y_n),
\end{equation}
where we have set the five-dimensional Newton constant to one, for
simplicity. The delta-functions in this equation are covariant with
respect to the fifth dimension, that is, they include a factor of
$1/\sqrt{-g_{55}}$. Furthermore, $T_{\a\b}$ is the bulk stress-energy
tensor induced by fields that propagate in the full five-dimensional
space time. The brane stress-energy tensors $T^{(n)}_{\a\b}$, on the 
other hand, originate from fields that are confined to the branes at
the orbifold fix points.

In order to proceed further, we need to specify these stress-energy
tensors. Two requirements should be taken into account when doing
this. First, one should use the fact that the background has a
maximally symmetric three-dimensional space. Secondly, the
brane stress-energy tensors should be restricted in a way that is
consistent with their nature of representing fields on the branes.
This latter requirement can be most easily implemented by using the
constraints that follow from the Einstein equation~(\ref{fieldeq})
itself. Concretely, the delta-functions on the right-hand side of this
equation have to be matched by corresponding delta-functions that
appear on the left-hand side. The appearance of these latter
delta-functions, however, is controlled by the structure of the
equations and the continuity assumptions about the metric discussed
above.

Let us see what this implies in detail. We start with the background
stress-energy. For the bulk, the most general form of this tensor
consistent with the three-dimensional maximal symmetry is
\bb
 {T^\a}_\b = \left(\begin{array}{ccc}
             \r&0&-r\\0&-p{\d^i}_j&0\\r&0&-q
             \end{array}\right)\; .
 \label{T}
\ee
In particular, we note that the $05$ component can be
non-vanishing. This possibility is, in fact, already realized for the
simple case of a bulk scalar field that depends on $t$ and $y$. As far
as the symmetry of the background metric is concerned, the background brane
stress-energy tensors should have the same structure as~(\ref{T}).
However, as we will see in a moment, there are two more requirements
that follow from the equations of motion, namely that the $55$ and
the $05$ components vanish. As a result, the background stress-energy 
on the branes has the form
\bb
 {T^{(n)\a}}_\b = \left(\begin{array}{ccc}\r^{(n)}&0&0\\
                  0&-p^{(n)}{\d^i}_j&0\\
                  0&0&0\end{array}\right)\; .
 \label{Tn}
\ee
Let us now proceed to the perturbed stress-energy tensors. Since we
are focusing on scalar perturbations we write the most general
perturbation of the background bulk tensor~(\ref{T}) that can be
expressed in terms of scalars on the maximally symmetric subspace.
This leads to
\bb
 \d {T^\a}_\b = \left(\begin{array}{ccc} \d\r&-(\r +p)b^{-2}v_{|j}&-\d r\\
              (\r +p)v^{|i}&-\d p\,{\d^i}_j+{\s^{|i}}_{|j}&-u^{|i}\\
              \d r+2r(\f +\G )-(\r +q)W&-b^{-2}u_{|j}&
              -\d q\end{array}\right)
 \label{dT}
\ee
where $v$ and $u$ are two potentials for ``velocity'' fields and $\s$,
satisfying ${\s^{|i}}_{|i}=0$, determines the anisotropic stress.
The perturbed brane stress-energy tensors should have the same
structure. However, as we will see below, the
equations of motion impose further constraints implying vanishing
$55$ and $5i$ components as well as vanishing anisotropic stress.
Therefore, the brane stress-energy perturbations are given by
\bb
 \d {T^{(n)\a}}_\b = \left(\begin{array}{ccc}
       \d\r^{(n)}&-(\r^{(n)}+p^{(n)})b^{-2}v^{(n)}_{|j}&-\d r^{(n)}\\
       (\r^{(n)}+p^{(n)})v^{(n)|i}&-\d p^{(n)}{\d^i}_j&0\\
       \d r^{(n)}-\r^{(n)}W&0&0\end{array}\right)\; .
 \label{dTn}
\ee
We would like to present the equations of motion based on the metric
(\ref{stp}) and on the above stress-energy tensors that follow
from the Einstein equation~(\ref{fieldeq}). However, in writing the
metric~(\ref{stp}) two gauge choices where involved and it is not, a
priori, clear that these choices can be made while, at the same time,
keeping the branes at $y=\text{const}$ hypersurfaces as we have conveniently
assumed in writing eq.~(\ref{fieldeq}). As the first choice, we
decided to work with the two-dimensional metric $\g_{ab}$ in
conformal gauge. Fortunately, this can be achieved while keeping the
branes at hypersurfaces $y=\text{const}$~[\ref{Lukas3}]. In addition, we have
chosen the generalized longitudinal gauge~(\ref{lg}) for the scalar
perturbations. A brane, described by $y=y_n$ before the gauge
transformation that leads to longitudinal gauge, is described by
$y=\tilde{y}-\x^5(y)=y_n$ after this gauge transformation, where
$\tilde{y}$ is the transformed $y$ coordinate. However,
since $\x^5(y_n)=0$, as discussed above, this equation is solved by
$\tilde{y}=y_n$ to linear order. We conclude that, in the new
coordinates that lead to the generalized longitudinal gauge, the brane
location is unchanged to the relevant linear order. In summary,
therefore, using the Einstein equation~(\ref{fieldeq}) with the branes
described by $y=y_n$ does not restrict the generality of our results
given the gauge choices that we have made.

The background equations following from~(\ref{fieldeq}) have been
given in ref.~[\ref{Lukas3}] for the case of stress-energy induced by
scalar fields and in ref.~[\ref{Binetruy}] for the case of ideal fluids.
For completeness and in order to incorporate some of
the generalizations that we have made (such as the inclusion of a
$05$ component of the bulk stress-energy) we will nevertheless
present these equations here. We find
\bba
 a^2b^2{G^0}_0 &\equiv& 3\left[2\frac{\dot{a}^2}{a^2}+\frac{\dot{a}\dot{b}}{ab}
                   -\frac{a''}{a}+\frac{a' b'}{ab}+kb^2\right]
                = a^2b^2\left[\r +\sum_{n=1}^2\r^{(n)}
                \bar{\d}(y-y_n)\right]\label{G00}\\
 a^2b^2{G^5}_5 &\equiv&3\left[\frac{\ddot{a}}{a}-\frac{\dot{a}\dot{b}}{ab}
                 -2\frac{{a'}^2}{a^2}-\frac{a' b'}{ab}+kb^2\right]
                 =-a^2b^2q \label{G55}\\
 a^2b^2{G^0}_5&\equiv&3\left[ -\frac{\dot{a}'}{a}+2\frac{\dot{a}a'}{a^2}
                 +\frac{\dot{a}b'}{ab}+\frac{a'\dot{b}}{ab}\right]
                 = -a^2b^2r \label{G05}\\
 a^2b^2{G^i}_j&\equiv&\left[ 3\frac{\ddot{a}}{a}+\frac{\ddot{b}}{b}-
                 \frac{\dot{b}^2}{b^2}-3\frac{a''}{a}-\frac{b''}{b}
                 +\frac{{b'}^2}{b^2}+kb^2\right]{\d^i}_j
                 =-a^2b^2\left[p+\sum_{n=1}^2p^{(n)}
                 \bar{\d}(y-y_n)\right]{\d^i}_j\label{Gij}
\eea
Here we have defined the delta-function $\bar{\d}$ which incorporates
a factor $1/ab$. Based on these equations we can now justify the
vanishing of the $55$ and $05$ components in the brane
stress-energy~(\ref{Tn}). 
Such components, if non-vanishing, would appear
on the right-hand sides of~(\ref{G55}) and (\ref{G05}) multiplied with
delta-functions. We should, therefore, have corresponding
delta-function terms on the left-hand sides of these equations. Since
the scale factors $a$ and $b$ are assumed to be continuous, delta
functions can only appear from second derivatives of these quantities
with respect to $y$. However, there are no such terms in
eq.~(\ref{G55}) and (\ref{G05}). Hence, we conclude that the 
$55$ and $05$ components in eq.~(\ref{Tn}) must vanish.

For the perturbations, we find to linear order
\begin{eqnarray}\label{einstein00}
(ab)^2 \delta G^0_{~0} &\equiv&  
     3\left[ 2 \frac{a'b'}{ab} - 2\frac{a''}{a}  - \frac{a'}{a}
      \frac{\partial}{\partial y}- \frac{\dot a}{a} \frac{\partial}
      {\partial t} \right] \Gamma -3\left[ \frac{\dot a'}{a} +
      2\frac{a' \dot a}{a^2} +\frac{\dot a}{a} \frac{\partial}
      {\partial y}\right] W -6\left[ 2\frac{\dot a^2}{a^2} +
      \frac{\dot a \dot b}{ab}\right]\phi\nonumber \\
&&\quad + 3 \left[ 3\frac{a'}{a}\frac{\partial}{\partial y}
    - \frac{b'}{b}\frac{\partial}{\partial y}
    -3\frac{\dot a}{a}\frac{\partial}{\partial t} 
    - \frac{\dot b}{b}\frac{\partial}{\partial t}+2kb^2\right]\psi 
    +b^2\left(2 \psi + \Gamma \right)^{|i}_{~|i} + 3 \psi''\nonumber\\
&=&a^2b^2\left\{ \d\r +\sum_{n=1}^2(\d\r^{(n)}+\G\r^{(n)})
   \bar{\d}(y-y_n)\right\}
\end{eqnarray}

\begin{eqnarray}\label{einstein55}
(ab)^2 \delta G^5_{~5} &\equiv& -6\left[ 2\frac{{a'}^2}{a^2} 
      + \frac{a'b'}{ab}\right]\Gamma 
      - 3\left[ \frac{\dot{a'}}{a} + 2\frac{a' \dot{a}}{a^2} +
      \frac{a'}{a}\frac{\partial}{\partial t} \right]W
      + 3\left[ 2\frac{\dot{a}\dot{b}}{ab} - 2\frac{\ddot{a}}{a} 
      - \frac{a'}{a}\frac{\partial}{\partial y} 
      - \frac{\dot a}{a}\frac{\partial}{\partial t}\right]\phi \nonumber\\
&&\quad +3\left[ 3\frac{a'}{a}\frac{\partial}{\partial y} + 
      \frac{b'}{b}\frac{\partial}{\partial y}-3\frac{\dot a}{a}
      \frac{\partial}{\partial t}+\frac{\dot b}{b}\frac{\partial}
      {\partial t}+2kb^2\right]\psi + b^2
      \left(2\psi - \phi\right)^{|i}_{~|i} - 3\ddot \psi\nonumber\\
&=&-a^2b^2\d q
\end{eqnarray}

\begin{eqnarray}\label{einstein05}
(ab)^2 \delta G^0_{~5} &\equiv&  3\left[\frac{a''}{a} - 
     2\frac{a'b'}{ab} - 2\frac{a'^2}{a^2} \right]W 
     +3\left[ 2\frac{\dot a'}{a} - 2\frac{\dot a b'}{ab} 
     -2\frac{a' \dot b}{ab} - 4 \frac{a' \dot a}{a^2} 
     +\frac{\dot a}{a}\frac{\partial}{\partial y}\right]\phi \nonumber \\
&&\quad +3\left[\frac{\partial^2}{\partial t \partial y} -
     \frac{b'}{b}\frac{\partial}{\partial t} 
     -\frac{\dot b}{b}\frac{\partial}{\partial y}\right]\psi
     - \frac{b^2}{2}{W^{|i}}_{|i}-3\frac{a'}{a}\dot\Gamma\nonumber\\
&=&a^2b^2\left\{ -\d r-\sum_{n=1}^2\d r^{(n)}\bar{\d}(y-y_n)\right\}
\end{eqnarray}

\begin{eqnarray}\label{einstein0i}
(ab)^2\delta G^0_{~i} &\equiv& \left\{\left[\frac{3}{2}\frac{a'}{a} 
     + \frac{b'}{b} + \frac{1}{2}\frac{\partial}{\partial y} \right] W 
     +\left[3\frac{\dot a}{a} + \frac{\dot b}{b}\right]\phi 
     +\left[ \frac{\dot b}{b}+\frac{\partial}{\partial t}\right]\Gamma
     + 2 \dot \psi \right\}_{|i}\nonumber\\
&=& a^2\left\{ -(\r +p)v-\sum_{n=1}^2(\r^{(n)}+p^{(n)})v^{(n)}
    \bar{\d}(y-y_n)\right\}_{|i}
\end{eqnarray}

\begin{eqnarray}\label{einsteini5}
(ab)^2\delta G^5_{~i} &\equiv& \left\{\left[3\frac{a'}{a} +
      \frac{b'}{b}\right] \Gamma+\left[ \frac{b'}{b} +
      \frac{\partial}{\partial y} \right]\phi 
      +\left[ \frac{3}{2}\frac{\dot a}{a} + \frac{\dot b}{b} 
      +\frac{1}{2}\frac{\partial}{\partial t} \right] W - 2 \psi'
      \right\}_{|i}\nonumber\\
&=& -a^2u_{|i}
\end{eqnarray}

\begin{eqnarray}\label{einsteinij}
(ab)^2\delta G^i_{~j} &\equiv&\left\{\left[-6\frac{a''}{a} -
       2\frac{b''}{b} + 2\frac{b'^2}{b^2}-
       3\frac{a'}{a}\frac{\partial}{\partial y}
       - 3\frac{\dot a}{a}\frac{\partial}{\partial t} 
       - \frac{\dot b}{b}\frac{\partial}{\partial t} 
       - \frac{b'}{b}\frac{\partial}{\partial y} 
       - \frac{\partial^2}{\partial t^2}\right] \Gamma\right. \nonumber \\
&&\quad +\left[ 2\frac{b' \dot b}{b^2} - 2 \frac{\dot b'}{b}
       - 6 \frac{\dot a'}{a} -3\frac{a'}{a}\frac{\partial}{\partial t}
       - 3\frac{\dot a}{a}\frac{\partial}{\partial y} 
       -\frac{\dot b}{b}\frac{\partial}{\partial y}
       -\frac{b'}{b}\frac{\partial}{\partial t}
       -\frac{\partial^2}{\partial t \partial y}\right] W \nonumber\\
&&\quad +\left[ 2\frac{\dot b^2}{b^2} - 2 \frac{\ddot b}{b}
       - 6 \frac{\ddot a}{a} -3\frac{a'}{a}\frac{\partial}{\partial y}
       - 3\frac{\dot a}{a}\frac{\partial}{\partial t}
       - \frac{\dot b}{b}\frac{\partial}{\partial t}
       - \frac{b'}{b}\frac{\partial}{\partial y} 
       - \frac{\partial^2}{\partial y^2}\right] \phi\nonumber \\
&&\quad\left. + \left[ 6 \frac{a'}{a}\frac{\partial}{\partial y}
       - 6\frac{\dot a}{a}\frac{\partial}{\partial t} 
       + 2\frac{\partial^2}{\partial y^2}
       - 2\frac{\partial^2}{\partial t^2}\right]\psi 
       + 2b^2(\psi -\phi+\G )^{|k}_{~|k}\right\}{\d^{i}}_j\nonumber\\
&&\quad -b^2(\psi -\phi +\G )^{|i}_{~|j}\nonumber\\
&=& a^2b^2\left\{ -\d p\,{\d^i}_j+{\s^{|i}}_{|j}
    -\sum_{n=1}^2(\d p^{(n)}+\G p^{(n)})\,
    {\d^i}_j\,\bar{\d}(y-y_n)\right\}
\end{eqnarray}
Given those results, we can now return to the question of why the
perturbations of the brane stress-energy tensors must have the specific
form~(\ref{dTn}). We recall that all quantities in our
metric~(\ref{stp}) are even except the off-diagonal perturbation $W$
which is odd under the $Z_2$ symmetry. From our continuity
assumptions, delta-function terms in the perturbed Einstein tensor
can, therefore, arise from first derivatives of $W$ with respect to
$y$ and second derivatives with respect to $y$ of all other
quantities. Inspection of the above equations shows that such terms
are absent in the $5i$ and $55$ components of the perturbed Einstein tensor.
Consequently, the corresponding components in the perturbed brane
stress-energy should vanish. Furthermore, all terms in $\d {G^i}_j$
that could potentially lead to delta-functions are proportional
to ${\d^i}_j$. This implies that the anisotropic stress on the brane,
which would contribute to the traceless part of
eq.~(\ref{einsteinij}), must vanish. As a result, the traceless part
of eq.~(\ref{einsteinij})
\bb
(\psi -\phi +\G )^{|i}_{~|j}-\frac{1}{3}(\psi -\phi +\G )^{|k}_{~|k}
  {\d^i}_j = -a^2{\s^{|i}}_{|j}
\ee
only involves the bulk anisotropic stress as a source term. If the
bulk anisotropic stress vanishes as well, as, for example, is the case
for a perfect fluid in the bulk, one concludes that
\bb
 \psi -\phi +\G = 0\; .
\ee
The quantity $\psi -\phi +\G$ is the analog of the four-dimensional
quantity $\psi_4 -\phi_4$ that also vanishes in the absence of
anisotropic stress. This correspondence will be made more explicit in
section V.

\section{Density fluctuations on the brane}

A systematic study of density fluctuation in five dimensions
requires solving the full set of five-dimensional equations of motion
given in the previous section. However, for specific questions it
might be useful to have some information about the metric restricted
to the brane. For example, it is this restricted metric that is felt
by matter which is confined to the brane. In this section, we are
going to derive such equations on the brane starting from the general
equations of motion above.

For a $Z_2$ even field the meaning of its value on the brane is quite
clear. A $Z_2$ odd field may jump across the brane so its value may
have a sign ambiguity. To simplify the notation, we define the value
of an odd field on the brane as the one that is approached from within
the interval $y\in [0,R]$. This is precisely the boundary value of the
field as viewed in the boundary picture and, at the same time
represents one half of the jump at the fix point. We also recall that
the scale factors $a$, $b$ and the perturbations $\phi$, $\psi$ and
$\G$ are even and hence continuous while the perturbation $W$ is odd
and may jump across the fix points. 

Let us start with the background equations of
motion~[\ref{Lukas3},\ref{Binetruy}].
We have already mentioned that the delta-function sources in
eq.~(\ref{G00}) and (\ref{Gij}) have to be matched by the terms
containing second $y$ derivatives of the scale factors $a$ and $b$.
This leads to
\bb\label{[a,b]}
{a'\over a}=\mp{1\over 6}ab\r^{(n)}\; ,\qquad
{b'\over b}=\pm\frac{1}{2}ab\left(\rho^{(n)}+p^{(n)}\right)\; .
\ee
These conditions, as well as the following ones, are valid at the brane
positions $y=y_n$ where the upper (lower) sign holds for the brane
$n=1$ ($n=2$). While the two other non-vanishing equations of motion
do not contain delta functions, they can still be restricted to the
brane. From the $05$ component~(\ref{G05}) we find
\bb\label{conservB}
\dot{\r}^{(n)}=-3{{\dot a}\over a}\left(\r^{(n)} +
                  p^{(n)}\right)\mp 2abr\; .
\ee
which represents to an energy conservation equation on the brane.
Note, however, that, in addition to intrinsic brane quantities, this
equation also involves the off-diagonal bulk stress-energy component
$r$. This reflects the simple fact that the branes are not isolated
systems but can exchange energy with the bulk. Finally, we should
consider the $55$ component~(\ref{G55}). Restricted to the branes it results in
an equation of motion for the values of the scale factors $a$ and $b$
on the brane given by
\bb
 \frac{\ddot{a}}{a}-\frac{\dot{a}\dot{b}}{ab}+kb^2
  = -{a^2b^2\over 3}\left[\frac{1}{12}\r^{(n)}\left(\r^{(n)}+3p^{(n)}\right)
    +q\right]\; .
\ee

\medskip

An analogous procedure can now be applied to the perturbed
equations. Observe that only the components
$\delta G^0_{~0}$, $\delta G^i_{~i}$, $\delta G^0_{~5}$ and
$\delta G^0_{~i}$ contain explicit delta-function terms. They should
be matched by terms containing first $y$ derivatives of $W$ and second $y$
derivatives of all other quantities. This leads to
\bba
 \psi ' &=& {{\dot a}\over a}W\pm {1\over 6}ab\left(\delta\r^{(n)}
            -\Gamma\rho^{(n)}\right) \label{jump1}\\
 \phi ' &=& -\left({{\dot a}\over a}+{{\dot b}\over b}+{\partial\over
            \partial t}\right)W\pm {1\over 3}ab\left(\delta\r^{(n)}
            -\Gamma\r^{(n)}\right)\pm\frac{1}{2}
            ab\left(\delta p^{(n)} - \Gamma p^{(n)}\right)\label{jump2}\\
 W &=& \mp\frac{a}{b}\left(\r^{(n)}+p^{(n)}\right)
        v^{(n)}\label{jump3} \\
 W &=& \frac{\d r^{(n)}}{\r^{(n)}} \label{jump4}
\eea
Interestingly, the last equation implies that the component
$\d {T^{(n)5}}_0$ of the brane stress-energy perturbation vanishes, as
can be seen by comparison with eq.~(\ref{dTn}). The component
$\d {T^{(n)0}}_5=-\delta r^{(n)}$, however, is non-zero and is, from
eq.~(\ref{jump3}), (\ref{jump4}) determined by
\bb
 \d r^{(n)} =
   \mp\frac{a}{b}\r^{(n)}\left(\r^{(n)}+p^{(n)}\right)v^{(n)}\; .
\ee
We have, therefore, found an important additional constraint on the perturbed
brane energy-momentum tensor~(\ref{dTn}). The quantity $\d r^{(n)}$
is, in fact, uniquely fixed by the other components. For vacuum
energy with $p^{(n)}=-\r^{(n)}$ on the branes $\d r^{(n)}$ is zero,
but it is generally non-vanishing otherwise. This is,
perhaps, somewhat surprising since one could have expected that a purely
four-dimensional stress-energy tensor on the brane (with all $5$ components
vanishing) should be allowed. Here we see that this is generally not
the case.

Next, we deal with the odd components $05$ and $5i$ of the
perturbed equations of motion given in~(\ref{einstein05}) and
(\ref{einsteini5}). Restriction to the branes leads us, after some
algebra, to
\bba
{\dot \delta}^{(n)} &=& -\left( 1 + w^{(n)} \right)\left({v^{(n)|i}}_{|i}
              - 3 \dot\psi\right) - 3\frac{\dot a}{a}\left(
              \frac{\delta p^{(n)}}{\delta \rho^{(n)}} -
              w^{(n)}\right)\delta^{(n)} \nonumber\\
            && - 2(1+w^{(n)})v^{(n)}a^2
              (\rho + q)\mp 2ab\left(\G+2\f-\d^{(n)}+\frac{\d
              r}{r}\right)\frac{r}{\r^{(n)}}\; \label{econ1}
\eea
and 
\bba
 \frac{{{\dot v}^{(n)}}_{~~~|i}}{b^2} &=&\left[ -\frac{\dot a}{a}\left( 1 - 3w^{(n)}
 \right)- \frac{\dot w^{(n)}}{1+w^{(n)}}+\frac{\dot b}{b} \right] \frac{{v^{(n)}}_{|i}}{b^2} 
 - \frac{\delta {p^{(n)}}_{|i}}{\delta \r^{(n)}}\frac{\d^{(n)}}{1+w^{(n)}} 
 - \phi_{|i} \nonumber\\ 
&& \mp 2 \frac{a}{b} \frac{1}{\rho^{(n)}} \left[ \frac{u}{1+w^{(n)}}
 - r v^{(n)} \right]_{|i} \; , \label{econ2}
\eea
where $w^{(n)} = p^{(n)}/ \r^{(n)}$ and $\d^{(n)} =\d\r^{(n)}/\r^{(n)}$
denotes the energy contrast on the branes. These equations represent
the conservation of energy and momentum for scalar perturbations,
including possible energy-momentum flow from the bulk onto the brane
or from the brane into the bulk. They should be compared with
the corresponding equations in four dimensions, eq.~(30) in
[\ref{bert}], taking into account that the variable $\theta$ of
[\ref{bert}] is related to the peculiar velocity $v$ via
$\theta = {v^{|i}}_{|i}$. Eq.~(\ref{econ1}) differs from the
four-dimensional result by the last two terms. They describe
the energy flux from the bulk onto the brane.
In eq.~(\ref{econ2}) we have two new terms with respect to the
four-dimensional equation. They describe momentum flux
between bulk and brane. The coupling between bulk gravity and
brane matter expressed via the above
equations is one of the main results of this paper. It shows that,
when considering scalar metric perturbations on the branes, the branes cannot
simply be viewed as an isolated system but have to be considered
together with the bulk environment. Practically, this implies that
frequently one cannot simply copy four-dimensional formulae
when dealing with physics on a brane that is embedded in a
higher-dimensional space. Finally, we restrict the $55$ component of
the equations of motion, eq.~(\ref{einstein55}), to the brane. We 
find the following evolution equation for the perturbations projected
onto the branes:
\bba
 b^2(2\psi -\f )^{|i}_{~|i}&-&3\ddot{\psi}-3\frac{\dot{a}}{a}\dot{\f}
 +3\left(\frac{\dot{b}}{b}-3\frac{\dot{a}}{a}\right)\dot{\psi}
 +6kb^2(\psi+\phi)+a^2b^2{\r^{(n)}}^2\left[\frac{1}{6}(1+3w^{(n)})\f
 \right.\nonumber\\
 &+&\left.\frac{\d q+2q\f}{{\r^{(n)}}^2}+\frac{1}{6}\left(
 1+\frac{3}{2}w^{(n)}\right)\d^{(n)}+\frac{\d p^{(n)}}{4\d\r^{(n)}}
 \d^{(n)}\pm\frac{a}{b}\frac{r}{\r^{(n)}}(1+w^{(n)})v^{(n)}\right] = 0\; .
\eea

\section{Matching to the four-dimensional effective theory}

In the previous subsection, we have derived a set of equations for
the metric on the branes, essentially by restricting the
five-dimensional equations of motion. These results may, for
example, be useful to analyze the evolution of matter that is confined
to the brane. However, the most important task is to extract
predictions for structure in the late universe from our formalism
of metric perturbations in brane-world theories. In this section,
we will explain the basic steps in this direction.

First, we should introduce the four-dimensional effective theory,
describing physics at low energy, that is associated to our
five-dimensional brane-world theory~(\ref{fieldeq}).
It is this four-dimensional theory that described the evolution
of the universe ``today'' and that is used for the interpretation
of observational results. Theoretical predictions, originating from
our brane-world theory, should therefore be formulated in terms of this
effective theory. The five- and the four-dimensional effective theory
are related by a vacuum state that constitutes a specific solution
of the five-dimensional theory and should respect the symmetries that
we expect to find in the four-dimensional theory. Specifically,
four-dimensional Lorentz invariance implies that the vacuum metric
should have the structure
\bb
ds^2=A^2(y)dx^\mu dx^\nu\eta _{\mu\nu}-B^2(y)dy^2\; .
\ee
The functions $A$ and $B$ have to be such that this metric solves the
five-dimensional theory in the vacuum configuration. For our
five-dimensional theory~(\ref{fieldeq}), the simplest possibility is
to have no stress energy in the vacuum which results in a flat vacuum
metric, $A,B=\text{const}$. In five-dimensional heterotic M-theory the
vacuum configuration is determined by certain potentials in the bulk
and on the branes that involve the dilaton~[\ref{losw1},\ref{losw2}].
In this case, $A$ and $B$ are non-trivial functions of $y$ and the
deviation from the flat vacuum metric is determined by the size of
the so called strong coupling expansion parameter.
The vacua proposed in ref.~[\ref{RS1},\ref{RS2}]
are based on a vacuum configuration with specific cosmological
constants in the bulk and on the branes and result in an exponential
function for $A$ in the coordinate system where $B=\text{const}$. Each one
of these different vacuum states is associated with its specific
low-energy theory. For the sake of simplicity and concreteness, we
will here focus on the first possibility, namely the flat vacuum.
This choice represents, at the same time, a good approximation for
five-dimensional heterotic M-theory in the case of a small
strong-coupling expansion parameter. The four-dimensional effective
theory describes the dynamics of the collective excitations of the
vacuum state. In our case, these excitation are given by a
four-dimensional metric $g_{4\m\n}$ and the modulus $\b$ describing
the size of the fifth dimension. The vacuum metric with these
collective modes put in has the structure
\bb
 d\bar{s}^2 = e^{-\b}g_{4\m\n}dx^\m dx^\n-e^{2\b}dy^2\; , \label{vacuum}
\ee
where $g_{4\m\n}$ and $\b$ are functions of $x^\m$.
As usual, the effective four-dimensional description is valid as long
as these functions are varying slowly enough. This is the case if all
four-dimensional momenta are much smaller than the mass of the first
Kaluza-Klein excitation around the vacuum state. In our case this mass
is given by $e^{-\b}/2R$. Let us, therefore, consider a
five-dimensional evolution that is approaching the vacuum
state~(\ref{vacuum}). Even though the five-dimensional metric is then
close to the vacuum metric it will still have small Kaluza-Klein
excitations that can be described in linear perturbation theory.
A useful way to extract the zero modes from such a five-dimensional
metric with small contributions from Kaluza-Klein modes is
to perform an average over the fifth dimension. Doing this
systematically leads to the following four-dimensional effective
theory associated to the brane-world theory~(\ref{fieldeq}) and the vacuum
state~(\ref{vacuum}):
\bba
 R_{4\m\n}-\frac{1}{2}g_{4\m\n}R_4 &=& \frac{3}{2}
   \left(\partial_\m\b\partial_\n\b -\frac{1}{2}g_{4\m\n}\partial\b^2\right)+
   T_{4\m\n} \\
 \nabla_4^2\b &=& J_4\; .\label{beom}
\eea
Four- and five-dimensional quantities are related by
\bba
 e^{2\b} &=& -<g_{55}> \label{c1}\\
 g_{4\m\n} &=& e^\b <g_{\m\n}> \\
 T_{4\m\n} &=& <T_{\m\n}>+\frac{1}{2
               Re^\b}\sum_{n=1}^2T_{\m\n}^{(n)} \\
 J_4 &=& \frac{2}{3}e^{-3\b}<T_{55}>+\frac{1}{3}g_4^{\m\n}T_{4\m\n}\;.
 \label{c4}
\eea
Here, $<\cdot >$ denotes the average over the fifth dimension. As
stated, this four-dimensional theory is a good description as long as
all momenta are small compared to $e^{-\b}/2R$, the mass of the first
Kaluza-Klein mode. The Kaluza-Klein modes have decoupled from the
above equations at linear order. However, due to the presence of the
branes, the Kaluza-Klein modes cannot strictly be set to zero but have
to be integrated out~[\ref{low4},\ref{low10}]. This leads to higher order
corrections to the above four-dimensional equations that are
suppressed by powers of the four-dimensional Planck scale and that we
have neglected. To be consistent with this approximation, the average
$<\cdot >$ that projects out the Kaluza-Klein excitations should be
considered meaningful only at the linearized level in Kaluza-Klein
excitations.

\medskip

We would now like to apply the above general correspondence to our
formalism for metric fluctuations. To do this, we need to assume
a five-dimensional background solution that, for late time, approaches
the vacuum configuration. Formulated in a four-dimensional language,
this is the case if the four-dimensional Hubble parameter
$H_4=\dot{a}_4/a_4$ and $\dot{\b}$ are small compared to $e^{-\b}/2
R$. Furthermore, it is helpful to assume that the average $<\g_{00}>$
approaches one in this limit. This can always be achieved by a
redefinition of time. We would like to explicitly work out the correspondence
for the scalar sector in longitudinal gauge that we have
focused on in this paper. The generalization to include vector and
tensor perturbations is straightforward.
Concretely, we apply the correspondence~(\ref{c1})--(\ref{c4})
to the five-dimensional quantities~(\ref{stp}), (\ref{T}), (\ref{Tn}),
(\ref{dT}) and (\ref{dTn}) matching onto the four-dimensional
quantities specified in (\ref{stp4}), (\ref{T4}) and (\ref{dT4}).
Furthermore, we need to decompose the four-dimensional modulus $\b$ as
\bb
 \b = \chi-\G_4\; .
\ee
where $\c =\c (t)$ is the time-dependent background and $\G_4
=\G_4(t,x^i)$ is the perturbation. The matching of background
quantities leads to
\bba
 e^{2\c} &=& <a^2b^2> \\
 a_4^2 &=& e^\c <a^2> \\
 \r_4 &=& e^{-\c}<\r >+\frac{1}{2 Re^{2\c}}\sum_{n=1}^2\r^{(n)} \\
 p_4 &=& e^{-\c}<p>+\frac{1}{2 Re^{2\c}}\sum_{n=1}^2p^{(n)}\; .
\eea
It is interesting to explicitly compute the background current
$J_4$ in the equation of motion~(\ref{beom}) for the modulus $\c$.
It is given by
\bb
 J_4 = \frac{1}{3}\left[\r_4-3p_4+2e^{-\c}<q>\right]\; .
\ee
With the above expression for $\r_4$ and $p_4$, this implies that
the modulus $\c$ has a runaway potential leading to a growing size of
the fifth dimension. Therefore, the theory, as stands, will not stay
in the range of validity of the four-dimensional effective theory.
As is well-known, it needs further stabilization of the modulus
$\b$ by means of a potential. In the context of string- or M-theory
one expects such a potential to be generated by non-perturbative effects.

The correspondence for the scalar perturbations reads
\bba
 \G_4 &=& <\G > \\
 \f_4 &=& <\f -\G/2> \\
 \psi_4 &=& <\psi +\G/2> \\
 \d\r_4 &=& e^{-\c}<\d\r +\G\r >+\frac{1}{2Re^{2\c}}
            \sum_{n=1}^2(\d\r^{(n)}+2\G_4\r^{(n)}) \\
 \d p_4 &=& e^{-\c}<\d p +\G p>+\frac{1}{2Re^{2\c}}
            \sum_{n=1}^2(\d p^{(n)}+2\G_4p^{(n)}) \\
 \s_4 &=& e^{-\c}<\s > \\
 v_4 &=& \frac{e^{-3\c}a_4^4}{\r_4+p_4}\left[<(\r +p)v>+
        \frac{1}{2 Re^{\b}}\sum_{n=1}^2(\r^{(n)}+p^{(n)})v^{(n)}
        \right]\; .
\eea
In particular, we conclude that
\bb
 \psi_4-\f_4 = <\psi -\f +\G >
\ee
Hence, the four- and five-dimensional quantities that measure the
presence of anisotropic stress are in direct correspondence to one
another as they should.

\section{Conclusion}

In this paper we have laid down a gauge-invariant formalism to
describe metric fluctuations in brane-world theories. This formalism
is a straightforward generalization of the well known formalism in
four dimensions. It categorizes the perturbations according to their
tensor properties with respect to the usual three-dimensional
maximally symmetric space rather than a higher-dimensional space
as one might have expected. This is a direct consequence of the
brane-world nature of the theory which generally leads to
cosmological backgrounds that are inhomogeneous in the additional
dimensions. We have introduced a generalized longitudinal gauge in order
to further study scalar perturbations. In the case of a
five-dimensional model on the orbifold $S^1/Z_2$, on which we have
focused, we have identified four scalar metric perturbations
$\phi$, $\psi$, $\G$ and $W$.
This has to be contrasted to the four-dimensional case where one only
has two such perturbations. We have presented the evolution equations for
these scalar perturbations which, mainly due to the dependence of the
background on the additional dimension, are significantly more
complicated than the corresponding four-dimensional equations.
It is those additional terms, related to the non-linearity of the
background in the additional coordinates, that encode possible new and
interesting information about the formation and evolution of perturbations.
Furthermore, given the gauge choices and assumptions about the 
continuity of the metric that we have made, we have determined 
the resulting most general form of the stress--energy on the brane.
In particular, we have found that the perturbed brane stress-energy
has to have vanishing anisotropic stress and that its $05$ component 
is non-zero. We have applied our formalism to calculate the matching
conditions (Israel conditions) for the five-dimensional metric
restricted to the branes. Among other results we have derived the
equations describing energy and momentum conservation for metric
perturbations on the brane. As is expected on physical grounds, they
illustrate that the brane cannot be viewed as an isolated object but is
subject to energy and momentum flow between the bulk and the brane. 
Finally, we have shown how the five-dimensional formalism
for metric fluctuations can be matched to the known four-dimensional
one in the limit where the brane-world theory has an effective
four-dimensional description. This allows one to extract
predictions for structure in the late universe originating from
brane-world theories. We hope to address this problem more
explicitly in a future publication.

\medskip\medskip

As this manuscript was prepared for submission, ref.~[\ref{muko}], [\ref{kodama}] and
[\ref{langlois00}] appeared which have some overlap with the present paper. 
Cosmological perturbations are also
discussed in [\ref{maartens}]. 

\begin{center}
{\bf Acknowledgments:}\\
\end{center}
C.~v.~d. Bruck is grateful to  E.~ Eyras, J.~Martin, C. Martins, 
H.~Reall and T.~Shiromizu, and in particular D. Langlois, R. Maartens 
and D. Wands for useful discussions. 
C.~v.~d. Bruck is supported by Nato/DAAD (at Brown) and Deutsche
Forschungsgemeinschaft (DFG, at Cambridge). Miquel Dorca is supported by
the {\em Fundaci\'on Ram\'on Areces}. The research was supported
in part (at Brown) by the U.S. Department of Energy under Contract 
DE-FG02-91ER40688, TASK A. A.~L. is supported by the European
Community under contract No.~FMRXCT 960090.

\references
\item \label{Horava} P. Ho\v rava and E. Witten, hep-th/9510209,
                     {\em Nucl. Phys.} {\bf B460} (1996) 506;
                     hep-th/9603142, {\em Nucl. Phys.}
                     {\bf B475} (1996) 96.
\item \label{witten} E. Witten, hep-th/9602070, {\em Nucl. Phys.}
                     {\bf B471} (1996) 135.
\item \label{hgaugino} P. Ho\v rava, hep-th/9608019, {\em Phys. Rev.}
                       {\bf D54} (1996) 7561.
\item \label{losw1} A. Lukas, B.~A. Ovrut, K.~S. Stelle and D. Waldram,
                    hep-th/9803235, {\em Phys. Rev.} {\bf D59} (1999)
                    086001.
\item \label{Dvali1} N. Arkani-Hamed, S. Dimopolous and G. Dvali,
                     hep-ph/9803315, {\em Phys. Lett.} {\bf B429}
                     (1998) 263.
\item \label{Dvali2} I. Antoniadis, N. Arkani-Hamed, S. Dimopolous
                     and G. Dvali, hep-ph/9804398, {\em Phys. Lett.}
                     {\bf B436} (1998) 257.
\item \label{kt} Z. Kakushadze and S.~H.~H. Tey, hep-th/9809147,
                 {\em Nucl. Phys.} {\bf B548} (1999) 180.
\item \label{lyk} J.~D. Lykken, hep-th/9603133, {\em Phys. Rev.}
                  {\bf D54} (1996) 3693.
\item \label{lowcosm} A. Lukas, B.~A. Ovrut and D. Waldram,
               hep-th/9806022, {\em Phys. Rev.} {\bf D60} (1999)
               086001.
\item \label{Dvali3} N. Arkani-Hamed, S. Dimopolous and G. Dvali,
                     hep-ph/9807344, {\em Phys. Rev.} {\bf D59}
                     (1999) 086004.
\item \label{real} H. Reall, hep-th/9809195, {\em Phys. Rev.}
                   {\bf D59} (1999) 103506.
\item \label{Lukas2} A. Lukas, B.~A. Ovrut and D. Waldram,
                     {\em Lectures given at the Advanced School on
                     Cosmology and Particle Physics}, June 1998,
                     hep-th/9812052 and references therein.
\item \label{Lukas3} A. Lukas, B.~A. Ovrut and D. Waldram,
                     hep-th/9902071, {\em Phys. Rev.} {\bf D61}
                     (2000).
\item \label{chamblin} H. Chamblin and H. Reall, hep-th/9903225,
                      {\em Nucl. Phys.} {\bf B562} (1999) 133.
\item \label{Binetruy} P. Binetruy, C. Deffayet and D. Langlois,
                       hep-th/9905012, {\em Nucl. Phys.} {\bf B565}
                       (2000) 269.
\item \label{Kaloper} N. Kaloper, hep-th/9905210, {\em Phys. Rev.}
                      {\bf D60} (1999) 123506.
\item \label{Csaki} C. Csaki, M. Graesser, C. Kolda and J. Terning,
                    hep-ph/9906513, {\em Phys. Lett.} {\bf B462}
                    (1999) 34.
\item \label{Cline} J. Cline, C. Grojean and G. Servant, hep-ph/9906523,
                    {\em Phys. Rev. Lett.} {\bf 83} (1999) 4245.
\item \label{freese} D.~J.~H. Chung and K. Freese, hep-ph/9906542 
                     {\em Phys. Rev.} {\bf D61} (2000) 023511.
\item \label{steinhardt} P.~J. Steinhardt, hep-th/9907080,
                        {\em Phys. Lett.} {\bf B462} (1999) 41.
\item \label{Kim} H.~B. Kim and H.~D. Kim, hep-th/9909053,
                  {\em Phys. Rev.} {\bf D61} (2000) 064003.
\item \label{ellwanger2} U. Ellwanger, hep-th/9909103,
                         {\em Phys. Lett.} {\bf B473} (2000) 233.
\item \label{dewolfe} O. De Wolfe, D. Freedman, S. Gubser and
                      A. Karch, {\em Modeling the Fifth Dimension with
                      Scalars and Gravity}, hep-th/9909134.
\item \label{flanagan} E.~E. Flanagan, S.-H. Tye and I. Wasserman, {\em
                       A Cosmology of the Brane World},
                       hep-ph/9909373.
\item \label{Kanti} P. Kanti, I. I. Kogan, K. A. Olive and M. Pospelov,
                    hep-ph/9909481, {\em Phys. Lett.} {\bf B468}
                    (1999) 31.
\item \label{maeda} T. Shiromizu, K.-I. Maeda and M. Sasaki,
                    {\em The Einstein Equation on the 3-Brane World},
                    gr-qc/9910076.
\item \label{Binetruy2} P. Binetruy, C. Deffayet, U. Ellwanger
                        and D. Langlois, hep-th/9910219,
                        {\em Phys. Lett.} {\bf B477} (2000) 285.
\item \label{vollick} D. Vollick, {\em Cosmology on a Three-Brane},
                      hep-th/9911181.
\item \label{Csaki2} C. Csaki, M. Graesser, L. Randall and
                     J. Terning, {\em Cosmology of Brane Models with
                     Radion Stabilization}, hep-th/9911406.
\item \label{ida} D. Ida, {\em Brane World Cosmology}, gr-qc/9912002.
\item \label{Kanti2} P. Kanti, I. I. Kogan, K. A. Olive and
                     M. Pospelov, hep-th/9912266, {\em Phys. Rev.}
                     {\bf D61} (2000) 106004.
\item \label{wands} R. Maartens, D. Wands, B.~A. Bassett and
                    I. Heard, {\em Chaotic Inflation on the Brane},
                    hep-ph/9912464.
\item \label{Ellwanger} U. Ellwanger, {\em Cosmological Evolution
                        in Compactified Ho\v rava-Witten Theory
                        Induced by Matter on the Branes}, hep-th/0001126.
\item \label{Hawking} S.~W. Hawking, T. Hertog and H.~S. Reall, 
{\em Brane New World}, hep-th/0003052
\item \label{Mohapat} R.~N. Mohapatra, A. P\v erez-Lorenzana and C.~A. 
de S. Pires, {\em Cosmology of Brane-Bulk Models in Five Dimensions}, 
hep-ph/0003328  
\item \label{deruelle} N. Deruelle and T. Dole\v zel, {\em Brane 
versus shell cosmologies in Einstein and Einstein-Gauss-Bonnet
theories}, gr-qc/0004021  
\item \label{viss2} C. Barcelo and M. Visser, {\em Living on the Edge:
Cosmology on the Boundary of Anti-de Sitter Space}, hep-th/0004056 
\item \label{elpp} J. Ellis, Z. Lalak, S. Pokorski and W. Pokorski,
                   hep-ph/9805377, {\em Nucl. Phys.} {\bf B540} (1999) 149.
\item \label{losw2} A. Lukas, B.~A. Ovrut, K.~S. Stelle and D. Waldram,
                    hep-th/9806051, {\em Nucl. Phys.} {\bf B552} (1999)
                    246.
\item \label{visser} M. Visser, hep-th/9910093, Phys. Lett. {\bf
                     B159}, 22 (1985).
\item \label{RS1} L. Randall and R. Sundrum, hep-ph/9905221,
                  {\em Phys. Rev. Lett.} {\bf 83} (1999) 3370.
\item \label{RS2} L. Randall and R. Sundrum, hep-th/9906064,
                  {\em Phys. Rev. Lett.} {\bf 83} (1999) 4670.
\item \label{lykken} J. Lykken, L. Randall, {\em The Shape of Gravity},
                     hep-th/9908076.
\item \label{Bardeen} J. Bardeen, {\em Phys. Rev.} {\bf D22} (1980) 1882.
\item \label{Brandenberger} V.~F. Mukhanov, H.~A. Feldman and
                            R.~H. Brandenberger, {\em Phys. Rep.}
                            {\bf 215} (1992) 203. 
\item \label{Israel} W. Israel, {\em Nuovo Cim.}  {\bf B44} (1966) 1.
\item \label{fabris} J. C. Fabris and M. Sakellariadou,
                     {\em Class. \& Quantum Gravity} {\bf 14} (1997) 725.
\item \label{martin} Y. Kubyshin and  J. Martin, {\em Limits on
                     Kaluza-Klein Models from COBE Results},
                     gr-qc/9604031; {\em On Compatibility of the 
                     Kaluza-Klein Approach with the COBE Experiment},
                     gr-qc/9507010.
\item \label{abbott} R.~B. Abbott, B. Bednarz and S.~D. Ellis,
                     {\em Phys. Rev.} {\bf D33} (1986) 2147.
\item \label{dilaton} R. Burstein, M. Gasperini, M. Giovannini,
V.~F. Mukhanov and G. Veneziano, {\em Phys. Rev.} {\bf D51} (1995) 6744.
\item \label{bert} C.-P. Ma and E. Bertschinger, {\em Astrophys. Journ.}
                   {\bf 455} (1995) 7.
\item \label{low4} A. Lukas, B. A. Ovrut and D. Waldram, hep-th/9710208,
                   {\em Nucl. Phys.} {\bf B532} (1998) 43.
\item \label{low10} A. Lukas, B.~A. Ovrut and D. Waldram,
                    hep-th/9801087, {\em Nucl. Phys.} {\bf B540}
                    (1999) 230.
\item \label{muko} S. Mukohyama, {\em Gauge-Invariant Gravitational
                   Perturbations of Maximally Symmetric Space-Times},
                   hep-th/0004067.
\item \label{kodama} H. Kodama, A. Ishibashi and O. Seto, {\em 
Brane World Cosmology: Gauge Invariant Formalism for Perturbation}, hep-th/0004160.
\item \label{langlois00} D. Langlois, {\em Brane Cosmological Perturbations}, hep-th/0005025
\item \label{maartens} R. Maartens, {\em Cosmological Dynamics on the
Brane}, hep-th/0004166

\end{document}